\documentclass[12pt]{iopart}
\usepackage{epsf,graphicx}
\usepackage{epsfig}
\begin{document}
\bibliographystyle{unsrt}

\title[Centrality dependence of the $N(\Omega)/N(\phi)$ ratios and $\phi$ $v_{2}$]{Centrality dependence of 
the $N(\Omega)/N(\phi)$ ratios and $\phi$ $v_{2}$ - a test of thermalization in Au+Au collisions at RHIC}

\author{Sarah-Louise Blyth (for the STAR Collaboration)}

\address{Lawrence Berkeley National Laboratory, One Cyclotron Road, Berkeley, California, 94720, 
USA\footnote{Current address: Physics Department, University of Cape Town, Rondebosch, South Africa, 7708} }
\ead{slblyth@lbl.gov} 

\begin{abstract}
We present the centrality dependence of the $N(\Omega)/N(\phi)$ ratios and $\phi$ $v_{2}$ measured in
Au+Au collisions at $\sqrt{s_{NN}}=200$ GeV by the STAR experiment at RHIC. The results are 
compared to measurements of other identified particles and recombination model expectations
in order to gain insight into the partonic collectivity and possible thermalization of the produced medium.
%A recombination model based on coalescence of thermal $s$-quarks describes the central $N(\Omega)/N(\phi)$
%ratio up to $p_{T}\sim4$ GeV/$c$. The $\phi$-meson $v_{2}(p_{T})$ and $\langle v_{2}\rangle$ results may 
%indicate partonic collectivity of the medium which increases with increasing centrality.
\end{abstract}

%\maketitle

\section{Introduction}

The phenomenon of large baryon/meson ratios of $\sim1$, e.g. for $p/\pi$ and $\Lambda/K^{0}_{S}$, 
has been observed at 
intermediate $p_{T}$ (2-5 GeV/$c$) in Au+Au collisions at RHIC \cite{Abelev:2006jr, Adams:2006wk} and is 
known as the 
so-called ``baryon anomaly''. 
It has also been observed that these ratios increase with increasing centrality.
The observation of these large ratios prompted the description of particle production at intermediate 
$p_{T}$ at RHIC by various quark recombination models \cite{Molnar:2003ff,Fries:2003kq,Hwa:2002tu,Greco} which have 
been successful in 
qualitatively describing the $p_{T}$ dependence of the ratios. Measurements of the centrality dependence of the
multistrange $N(\Omega)/N(\phi)$ ($sss/s\bar{s}$) ratios can help to provide further insight into particle production
mechanisms at intermediate $p_{T}$ at RHIC.

The $\phi$-meson ($s\bar{s}$) is an ideal probe of the properties of the medium produced in nucleus-nucleus 
collisions at RHIC; with its assumed small interaction cross-section with non-strange hadrons~\cite{Shor:1984ui}
and its relatively long lifetime of 41 fm/$c$, which means it mostly decays outside the fireball, it can provide
a clean signal, undisturbed by subsequent hadronic interactions, from the early stage of the system's evolution. 
The $\phi$-meson can also be used to differentiate between mass-type and particle-type effects since it is a 
meson, but is comparable in mass to the proton and $\Lambda$ baryon.

Elliptic flow ($v_{2}$) is an observable which provides information from the early stages of the collision.
At RHIC, the $v_{2}(p_{T})$ of identified hadrons at low $p_{T}$ ($<2$ GeV/$c$) has been found to follow a 
mass ordering consistent with expectations from hydrodynamic models. At intermediate $p_{T}$ ($2<p_{T}<5$ GeV/$c$), 
identified particle $v_{2}(p_{T})$ has been observed to scale with quark 
number~\cite{Adams:2003am}. This is also consistent with the idea of particle production through quark coalescence 
and may shed light
on the deconfinement of the medium produced in the early stages of relativistic nucleus-nucleus collisions at RHIC.
In addition, the multi-strange baryons, $\Xi$ and $\Omega$, have $v_{2}$ values comparable to that of the lighter
particles which may be evidence for partonic collectivity of the produced medium~\cite{Adams:2005zg}. 
Measurements of the 
centrality dependence of the relatively heavy $\phi$-meson can help to differentiate between particle-type and
mass effects in $v_{2}(p_{T})$ and provide further insight into the collectivity of the medium.

\section{Experiment and Analysis}
In these proceedings we present data from Au+Au collisions at $\sqrt{s_{NN}}=200$ GeV measured by 
the STAR experiment at RHIC. Approximately 13.5 million minimum bias events and 19 million central-triggered 
events were analyzed to produce the presented particle ratios and $v_{2}$ results. The $\phi$-mesons were reconstructed 
%on a statistical basis
via their decay to two oppositely charged kaons identified by their $dE/dx$ energy loss in the STAR 
Time Projection Chamber (TPC). The invariant mass distribution was constructed using all
combinations of oppositely charged kaons per event and the uncorrelated background was estimated
using event-mixing~\cite{Adams:2004ux}. 
The resulting 
$\phi$ mass peak was then fitted with a Breit-Wigner function (to describe the signal) plus a straight line 
(to describe the residual background).
The $v_{2}(p_{T})$ results were obtained using the $v_{2}$ vs. mass method following~\cite{Borghini:2004ra}. 
The systematic errors in the $v_{2}$ include the difference in the results obtained using two different methods
and from extracting the raw yields using different fitting criteria.

\section{Results}

\begin{figure}[hbt!]
\center
\includegraphics[scale=0.6]{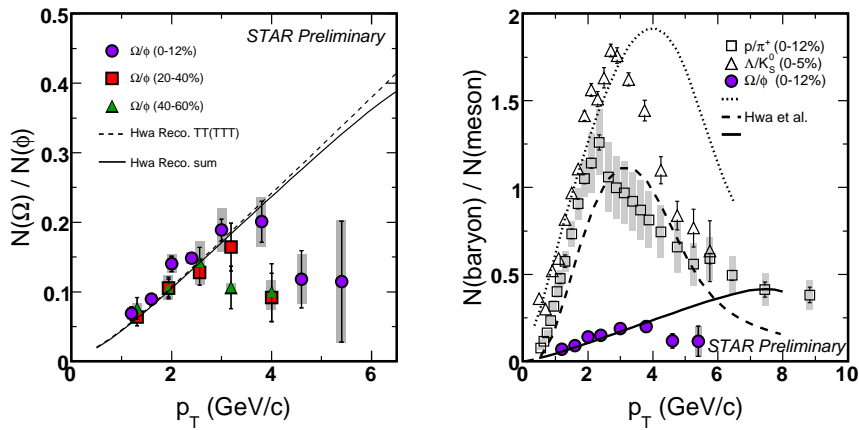}
\caption{\label{fig:OmPhiRatio} \textbf{Left panel:} Centrality dependence of the $N(\Omega)/N(\phi)$ ratios
vs. $p_{T}$. Recombination model expectations from~\cite{Hwa:2006vb} for central collisions
are shown by the dashed (thermal-thermal contribution) and solid lines (sum of all contributions).
\textbf{Right panel:} Central baryon/meson ratios for $\Lambda/K^{0}_{S}$ (triangles), $p/\pi$ (squares),
and $N(\Omega)/N(\phi)$ (circles) compared to recombination model expectations by Hwa et al.}
\end{figure}

In the left panel of figure~\ref{fig:OmPhiRatio} we present the centrality dependence of the $N(\Omega)/N(\phi)$ 
ratios vs. $p_{T}$.
The $\Omega$ spectra used to obtain the ratios were taken from~\cite{Adams:2005dq} for the 
central ratio and \cite{Adams:2006ke} for the 20-40\% and 40-60\% centrality ratios. 
The systematic and statistical errors in the ratios are dominated by the errors in the $\Omega$ spectra.
Also shown in figure~\ref{fig:OmPhiRatio} are recombination model calculations for central collisions, 
where it is expected that the dominant contribution to $\Omega$ and $\phi$ production 
is from coalescence of thermal $s$-quarks (TT contribution, dashed line)~\cite{Hwa:2006vb}.
For all centralities, at low $p_{T}$, the ratios increase monotonically while the
``turn-over'' points seem to shift towards higher $p_{T}$ with increasing centrality.
This may imply larger thermal $s$-quark contributions to multistrange hadron production in more central
Au+Au collisions at RHIC~\cite{Hwa:2006vb}.
The model describes the central data well up to $p_{T}\sim4$~GeV/$c$ after which it overpredicts the ratio.
The $p_{T}$ range 0-4 GeV/$c$ covers more than 95\% of the total yields for the $\phi$ and $\Omega$
and within the framework of the model~\cite{Hwa:2006vb}, the majority of the multistrange hadrons are formed
through the coalescence of thermal $s$-quarks in central Au+Au collisions at RHIC.

The right panel of figure~\ref{fig:OmPhiRatio} compares the central $N(\Omega)/N(\phi)$ ratio to that for 
$p/\pi^{+}$ and $\Lambda/K^{0}_{S}$ and includes recombination model expectations from 
Hwa et al.~\cite{Hwa:2006vb,Hwa:2004ng}. 
For all
cases as a function of $p_{T}$, the particle ratios increase, have a turning point at intermediate 
$p_{T}$ and then decrease. Also noticeable is a possible shift towards higher $p_{T}$ 
in the turning points of the ratios as a function of increasing strangeness content of the particles.
This may imply different production mechanisms for strangeness compared to the lighter $u$ and $d$ flavours.
For all the particle ratios, the recombination model expects the turning points at higher $p_{T}$ values than
observed in the data.

\begin{figure}[hbt!]
\center
\includegraphics[scale=0.55]{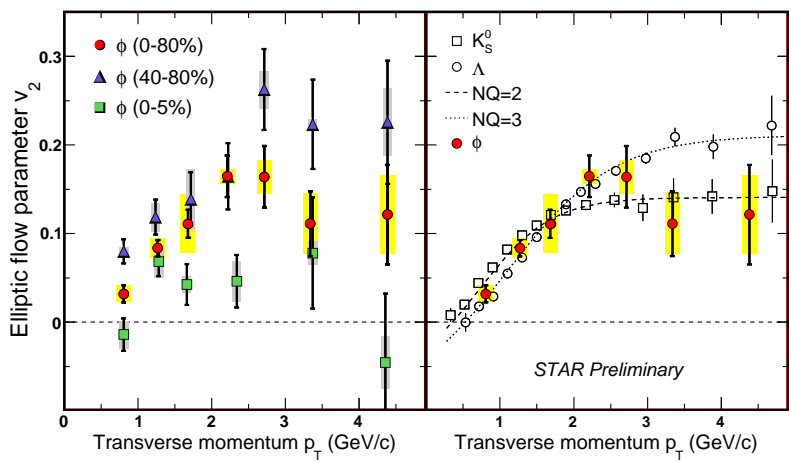}
\caption{\label{fig:v2} \textbf{Left panel:} Centrality dependence of $\phi$ $v_{2}(p_{T})$.
\textbf{Right panel:} Minimum bias $\phi$ $v_{2}(p_{T})$ (full circles) 
compared to $\Lambda$ (open circles) and $K^{0}_{S}$ (squares) and quark number (NQ) scaling 
parameterizations (lines).}
\end{figure}

In the left panel of figure~\ref{fig:v2}, we present the centrality dependence of the $\phi$ $v_{2}(p_{T})$.
The $v_{2}(p_{T})$ values increase with decreasing centrality i.e. with increasing eccentricity of the 
initial overlap region of the colliding nuclei. This observation is consistent with measurements of other
hadrons~\cite{Adams:2004bi}.

The right panel of figure~\ref{fig:v2} compares the $\phi$-meson minimum bias (0-80\%) $v_{2}(p_{T})$ 
to that of $\Lambda$ and $K^{0}_{S}$ from~\cite{Adams:2003am}. The dashed
and dash-dotted lines indicate parameterizations based on quark number (NQ) scaling for mesons and baryons 
respectively~\cite{Dong:2004ve}. For $p_{T}<2$ GeV/$c$, a mass scaling consistent with hydrodynamic expectations
is observed for the $\phi$, however higher precision data are needed to allow a final conclusion to be drawn.
At intermediate $p_{T}$ (2-5 GeV/$c$), the $\phi$ $v_{2}(p_{T})$ is consistent with that of $K_{S}^{0}$.
This particle type dependence, i.e. baryon/meson dependence of the elliptic flow may imply that the system
created in Au+Au collisions at RHIC is in a state of deconfinement before hadronization.

\section{Summary}
We have presented the centrality dependence of the $N(\Omega)/N(\phi)$ ratios and $\phi$-meson $v_{2}(p_{T})$
measured by the STAR experiment in Au+Au collisions at $\sqrt{s_{NN}}=200$ GeV.
The $\phi$ spectra in central collisions can be described by thermal-type distributions 
and a recombination model based on coalesence of thermal $s$-quarks~\cite{Hwa:2006vb} describes 
the central $N(\Omega)/N(\phi)$
ratio up to $p_{T}\sim4$ GeV/$c$, a range which covers 95\% of the yields of these particles. 
In addition, the $v_{2}(p_{T})$ of the $\phi$ is consistent with other mesons which implies partonic
collectivity of the system in the early stage of Au+Au collisions at RHIC.

%\begin{equation*}
%\frac{1}{M_{\rm a}}\left(\int^\infty_0{\rm d}
%\omega\;\frac{|S_o|^2}{N}\right)^{-1}\qquad\mbox{instead of}\qquad
%\frac{1}{M_{\rm a}}\biggl/\int^\infty_0{\rm d}
%\omega\;\frac{|S_o|^2}{N}.
%\end{equation*}

%\begin{table}
%\caption{\label{label}Table caption.}
%\begin{indented}
%\item[]\begin{tabular}{@{}llll}
%\br
%Head 1&Head 2&Head 3&Head 4\\
%\mr
%1.1&1.2&1.3&1.4\\
%2.1&2.2&2.3&2.4\\
%\br
%\end{tabular}
%\end{indented}
%\end{table}

\section*{References}
\bibliography{Bibliography}

\end{document}